\documentstyle[epsfig,aps,pra]{revtex}
\begin{document}

\newcommand{\ket}[1]{| #1 \rangle}
\newcommand{\bra}[1]{\langle #1 |}
\newcommand{\braket}[2]{\langle #1 | #2 \rangle}
\newcommand{\proj}[1]{| #1\rangle\!\langle #1 |}

\twocolumn[\hsize\textwidth\columnwidth\hsize\csname
@twocolumnfalse\endcsname

]

\centerline {\Large \bf Multi-Player Quantum Games} 
\bigskip

\centerline {\bf Simon C. Benjamin and Patrick M. Hayden}
\smallskip
\centerline {Centre for Quantum Computation, Clarendon}
\centerline {Laboratory, University of Oxford, OX1 3PU, UK.}
\bigskip

{\bf 
Recently the concept of quantum information has been introduced into game theory. Here
we present the first study of quantum games with more than two players. We discover that such games can possess a new form of
equilibrium strategy, one which has no analogue either 
in traditional games or even in two-player quantum games. In these `pure' coherent equilibria, entanglement shared among multiple players enables new kinds of
cooperative behavior: indeed it can act as a contract, in the sense that it prevents players from successfully betraying one-another.}

\bigskip

Game theory is a mature field of applied mathematics. It formalizes the conflict between
competing agents, and has found applications ranging from economics
through to biology
\cite{GTbook,bioNature}.  Quantum information is a young field of physics. At its heart is the realization
that information is ultimately a physical quantity, rather than a mathematical abstraction
\cite{physInfo}. It is known that various problems in this field can be usefully thought of as games. Quantum
cryptography, for example, is readily cast as a game between the individuals who wish to communicate, and
those who wish to eavesdrop \cite{qCrypt}. Quantum cloning has been thought of as a physicist playing a game
against nature \cite{clone}, and indeed even the measurement process itself may be thought of in these terms
\cite{deutschMeasure}. Furthermore, Meyer
\cite{MeyerNotes,MeyerPRL} has pointed out that the algorithms conceived for quantum computers may be
usefully thought of as games between classical and quantum agents. Against this background, it is natural to
seek a unified theory of games and quantum mechanics \cite{natureReview,ScienceNewsReview,SciAmReview}. Such
a theory might lend insight into biological and chemical processes occurring in the quantum regime; it would
certainly provide a fuller understanding of the physics of information \cite{physInfo}.

The fundamental unit of classical information is the {\em bit}. The
corresponding unit of quantum information is the `qubit' -- a general quantum superposition of `0' and `1', 
$\alpha_0\ket{0} + \alpha_1\ket{1}$. In multi-qubit systems, superposition gives rise to {\em
entanglement}: qubits are entangled if their states cannot be defined independently from one
another.
  Whereas a pair of classical
bits must be in one of the four states $\{00, 01, 10, 11 \}$, a pair of qubits can be in a
state, such as 
${1\over\sqrt{2}}(
\ket{0}\otimes\ket{0}+\ket{1}\otimes\ket{1}$), which cannot be factorized into two separate qubit
states. The interdependence remains even when the two qubits are far apart - this is the origin of
`non-local' effects in quantum mechanics. Although the effect cannot directly transfer information, it
has been identified as a crucial resource in quantum communication, quantum computation and
error-correction, and some forms of quantum cryptography \cite{physInfo}. Here we will see that when the
resources controlled by competing agents are entangled, they can cooperate to perfectly exploit their
environment (i.e. the `game'), and to prevent one-another from `defecting'.

Formally a {\em game}
involves of a number of agents or {\em players}, who are allowed a certain set of moves or
{\em actions}. The {\em payoff function}
$\$()$ specifies how the players will be rewarded after they have performed their actions. The
$i^{th}$ player's {\em strategy},
$s_i$, is her procedure for deciding which action to play, depending on her information. The {\em
strategy space},
$S=\{s_i\}$, is the set of strategies available to her.  A
{\em strategy profile} $s=(s_1,s_2,..,s_N)$ is an assignment of one strategy to each player. We will use the term {\em equilibrium} purely in its game theoretic sense, i.e. to refer to a
strategy profile with a degree of stability -- for example, in a Nash equilibrium no player can
improve her expected payoff by unilaterally changing her strategy. The study of equilibria is
fundamental in game theory \cite{GTbook}. 
The games we consider here are {\em static}: they are
played only once so that there is no history for the players to consider. Moreover, each player has
complete knowledge of the game's structure. Thus the set of allowed actions corresponds
directly to the space of deterministic strategies.  

Our procedure for quantizing games is a
generalization of the elegant scheme introduced by Eisert {\it et al.} \cite{EisertNotes,EisertPRL}. We
reason as follows. Game theory, being a branch of applied mathematics, defines games without reference to
the physical universe. However, quantum mechanics is a physical theory, and must be applied to a physical
system. We therefore begin by creating a physical model for the games of interest. A very natural
way to do this is by considering the flow of information, see Fig 1(a). 

\begin{figure}
\centerline{\epsfig{file=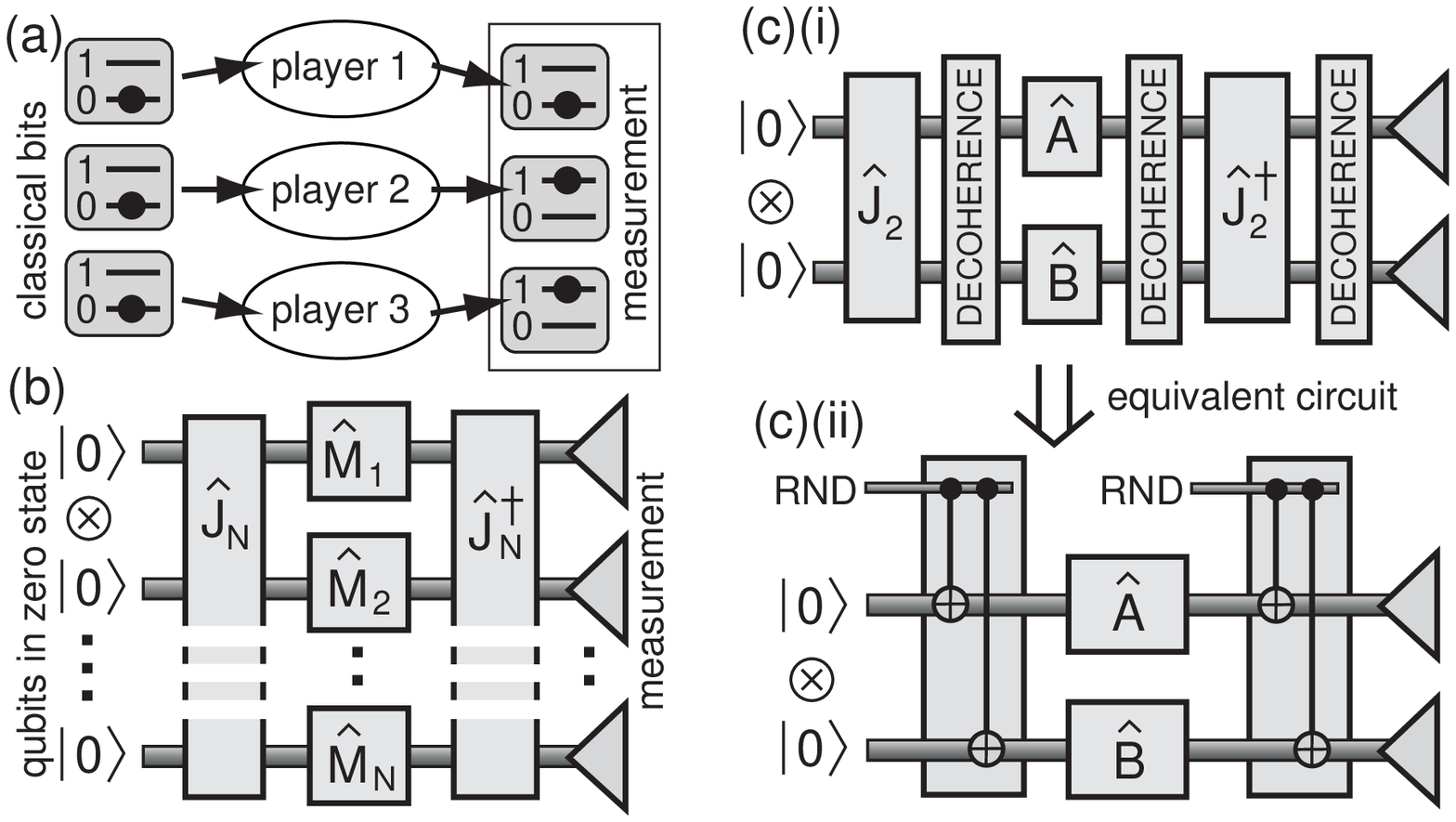,width=7.8cm}}
\vspace{0.2cm}
\caption{(a) a physical model for a game in which each
player has two possible actions: we send each player a classical 2-state system (a
bit) in the zero state. They locally manipulate their bit in whatever way they wish: under classical
physics their choices are really just to flip, or not to flip. They then return the bits for
measurement, from which the payoff is determined. (b) Our
$N$-player quantized game. Throughout this paper, `measurement' means measurement in the computational
basis, \{$\ket{0},\ket{1}$\}. (c) The effect of introducing total {\em decoherence} of the quantum
information.  RND denotes a random classical bit, the vertical lines denote CONTROL-NOT.}
\label{figure1}
\end{figure}

\noindent This classical physical model is then to be quantized. Our quantization procedure is the most
natural one that meets the following requirements: (a) The classical information carriers (bits) are to
be generalized to quantum systems (qubits). (b) These qubits are to be mutually entangled \cite{whyEnt}.
(c) The resulting game must be a {\em generalization} of the classical game: the identity operator
${\hat I}$ should {\em correspond} to `don't flip', and the bit-flipping operator ${\hat F}={\hat
\sigma}_x$ should {\em correspond} to `flip' \cite{whySigmaX}, in the sense that when all the players
restrict themselves to choosing from
$\{{\hat F},{\hat I}\}$, then the payoffs of the classical game are recovered. 
To simultaneously meet requirements (b) and (c), we employ a pair
of entangling gates as shown in Fig 1(b), and insist that ${\hat J}$ commutes
with any operator formed from ${\hat F}$ and ${\hat I}$ acting in the subspaces of different qubits. If
we restrict ourselves to unitary, maximally entangling gates \cite{whatIsMax}
that act symmetrically on ones and zeros, then we may specify ${\hat J}$ without loss of generality
\cite{stillGeneral}:
${\hat J}={1 \over \sqrt{2}}({\hat I}^{\otimes N}+i{\hat F}^{\otimes N})$. 

The representation in Fig. 1 (b) allows one to regard quantized games as simple quantum algorithms. The games we consider below could in fact be realized in a quantum computer possessing very few qubits (between one and three qubits per player, depending on the generality of the strategy space) \cite{longPaper}. NMR quantum computers already have sufficient qubits for this purpose.

In comparing the quantum and classical games, the choice of strategy space is fundamental. The
classical game is to be embedded in the quantum game, therefore the space should include playing the
`classical' actions
$\{ {\hat I}$,${\hat F}$\}, but in principle we could choose any superset
of this `classical' space. Previous studies have considered two-player games, and have employed strategy sets of limited generality. For example, in Ref \cite{MeyerPRL} Meyer explored the consequences of giving one player a full unitary strategy space whilst constraining the other to use only the `classical' space -- as one might expect, the quantum player dominates the game. Meyer has provided \cite{MeyerNotes} an interesting interpretation of such one-sided games wherein the players are a quantum computer and its operator. In a second approach \cite{EisertPRL}, Eisert {\em et al} permitted both players the same strategy set, but introduced an arbitrary constraint into that set \cite{comment}. This amounts to permitting a certain strategy $Q$ whilst forbidding the logical counter strategy -- as one might intuitively expect, $Q$ emerges as the ideal strategy. In contrast to these earlier approaches, throughout the present paper we allow all of our
players to perform any action on their qubits which is quantum mechanically possible. This includes adjoining
arbitrarily large ancillas, performing measurements and applying operations conditioned on the outcomes
of those measurements. We believe this to be the most natural generalization of our classical model, where the only
restrictions on the actions of the players were those
imposed by classical physics.  General
quantum operations are represented by trace-preserving, completely-positive maps, and we denote
the space of strategies corresponding to all such operations by $S_{TCP}$. 

In traditional game theory, there is a fundamental distinction between
so-called `pure' strategies, in which  players choose their actions deterministically, and `mixed'
strategies which can involve probabilistic choices. An important consequence of adopting a
general quantum model is that the players
can implement any probabilistic strategy entirely deterministically through the use of ancillary qubits.
For example, such qubits could function as a random number generator controlling the operations applied
to the primary qubit. Thus, all quantum strategies could be argued to be `deterministic'. Even so,
there is a subset of
$S_{TCP}$ that is in many ways analogous to the classical deterministic strategies, namely the set of all
strategies which correspond directly to a unitary action. Strategies from this subset, which we label
$S_U$, imply coherent manipulations of the local qubits, i.e. manipulations without
the addition of ancillary qubits.  Another way of identifying the set $S_U$ is that they are
precisely the strategies that do not destroy any of the entanglement introduced by the ${\hat J}$ gate
\cite{whyProbSU2destEnt}.  

In the multi-player games below,
we discover that equilibria for {\em all} of $S_{TCP}$ can consist of strategies drawn {\em only} from $S_U$.
We will refer to these special equilibria as {\em pure}, or {\em coherent}. They are
{\em fundamentally} quantum mechanical, in that they disappear when the quantum correlations
implicit in the entangled states are replaced with classical correlations, as in Fig. 1(c). In analogous
two-player games (where both players are permitted $S_{TCP}$) it is impossible for `pure'
equilibria
\cite{comment} to occur -- instead equilibria exist only when the players choose to degrade the
entanglement. Unsurprisingly therefore, those equilibria do persist in the Fig 1(c) variant.

Consider the classical $N$-player Minority Game \cite{minorityGame}. Here each player privately chooses
between two options, say `0' and `1'. The choices are then compared and the
player(s) who have made the minority decision are rewarded (by one point, say). If there is
an even split, or if all players have made the same choice, then there is no reward. The structure of
this game reflects many common social dilemmas, for example choosing a route in rush hour, choosing
which evening to visit an overcrowded bar, or trading in a financial market.
We can immediately quantize this game as discussed above and shown in Fig. 1.

We begin with $N=3$. Does quantization introduce new equilibria? Yes: for example, the players can coordinate their actions simply by measuring their qubits and exploiting the classical correlations \cite{longPaper}. However, such strategies are of limited interest in the present context, since they also function in the decoherent circuit of Fig 1(c). In fact, we can quickly prove that there are no novel `pure' strategies in this game: The most general pure strategy for player $i$ can be written $s_i=\alpha_i^A(\beta_i^A i\sigma_x + \beta_i^B i \sigma_y)+\alpha_i^B(\gamma_i^A I + \gamma_i^B i \sigma_z)$, where all the $\alpha$,$\beta$ and $\gamma$ coefficients are real and $(\alpha_i^A)^2+(\alpha_i^B)^2=(\beta_i^A)^2+(\beta_i^B)^2=(\gamma_i^A)^2+(\gamma_i^B)^2=1$. Then simply by applying the $J$ gates and deriving measurement probabilities in the standard fashion, we find that the $\beta$ and $\gamma$ terms disappear, yielding 

\noindent {\small PROB}(player 1 in minority)= $(\alpha_1^B\alpha_2^A\alpha_3^A)^2+(\alpha_1^A\alpha_2^B\alpha_3^B)^2$,

\noindent and similarly for players 2 and 3. But these are just the probabilities that occur in the classical game, when we identify the probability of player $i$ choosing to flip as $(\alpha^A_i)^2$.
Thus the extra parameters in the quantum strategies are of no significance, and the quantum game simply reduces to the classical variant.

Surprisingly, the situation is completely different in the
4-player Minority Game. Classically, the players have no better strategy than to choose
randomly between the `0' and `1' actions. The expected payoff for each player is
then one eighth of a point, i.e. the game only `pays out' half the time. But when we quantize the game, for the first time we discover fully coherent equilibria. One example \cite{manyNash} is the profile
$s=(a,a,a,a)$ where $a={1\over {\sqrt 2}}cos({\pi\over 16})(I+i\sigma_x)+{1\over {\sqrt 2}}sin({\pi\over 16})(i\sigma_y-i\sigma_z$). With these choices, the final state prior to
measurement is
$2^{-{2\over 2}}(|1000\rangle +|0100\rangle +|0010\rangle+|0001\rangle -|1110\rangle -|1101\rangle -|1011\rangle -|0111\rangle )$, i.e. an equal superposition
of eight states, two optimal for each player. Thus each player has expected
payoff ${1\over 4}$ -- twice the performance of the classical game and the logical maximum for a cooperative
solution. The reasoning below proves that the profile $s$ is a true Nash equilibrium: even though the players are allowed the full generality of $S_{TCP}$, no player can improve her expected payoff by unilaterally
defecting from $s$.

\begin{enumerate}
\item \label{proof1} 
Note that the Minority Game has the special property that the same expected payoffs result whether
or not we apply the second gate,
${\hat J}^\dagger$, prior to measurement. 
This can be seen by noting that
${\hat J}^\dagger$ transforms any basis vector $|abcd\rangle$ only within the sub-space spanned by 
\{$|abcd\rangle,|\bar{a}\bar{b}\bar{c}\bar{d}\rangle$\}, where 
$\bar{x} = NOT(x)$.  Since both $|abcd\rangle$ and
$|\bar{a}\bar{b}\bar{c}\bar{d}\rangle$ have the same payoff value, the expected payoff is 
left invariant by the $\hat{J}^\dagger$.

\item \label{proof2}
Because of (\ref{proof1}), we can focus attention on the state prior to ${\hat J}^\dagger$. This state
has the property that measurement of any three of the four qubits will yield one of the eight outcomes,
$(000)$, $(001)$, ...,$(111)$, with {\em equal} probability.  This must remain true regardless of
what local action was performed on the fourth qubit. Violation of this physical principle would mean
that entanglement could be used for faster than light information transfer, for example.

\item \label{proof3}
Six of these eight outcomes are {\em unwinnable} by the fourth player: if, for example, measurement of the
first three qubits yields $(001)$, then neither a `0' or a `1' will put the fourth player in the
minority. Thus, because of the equal weighting of the outcomes, her expected payoff cannot exceed
${1\over 4}$. {\em But this is just the payoff each player has with the originally proposed strategy
profile.} 
\end{enumerate}

Thus in moving from the $N=3$ to the $N=4$ player Minority game, a fundamentally new, non-classical equilibrium becomes available.
This equilibrium is optimal and fair: the game
always pays out the maximum amount \emph{and} the expected payoff for each of the players is the same. 
In the classical Minority Game, this \emph{can} be achieved, but only by sharing additional classical information
\cite{additClassRes}. We are therefore led to ask, are there games with
pure quantum equilibria whose performance {\em cannot} be matched classically {\em even}
in the presence of free communication? Surprisingly, the answer is yes. To 
demonstrate, we exploit the concept of `dominant' strategies.

A player has a
dominant strategy if this strategy yields a higher payoff than any alternative, {\em regardless} of
the strategies adopted by other players. A rational player will inevitably adopt such a strategy -- even when we allow free conversation with other
players (unless we introduce some kind of binding contract, which amounts to
switching to another payoff table entirely). Most games, including the Minority Game considered earlier, do not
possess dominant strategies.
If every player has a dominant strategy, then the game's inevitable outcome is the {\em
dominant-strategy equilibrium}. The famous Prisoner's Dilemma, shown in Fig 2(a),  has the
dominant-strategy equilibrium (`defect',`defect'). As noted above, no maximally entangled two-player
quantum game can have equilibria in the strategy space
$S_U$. Thus, quantization of Prisoner's Dilemma removes the dominant-strategy equilibrium
\cite{EisertNotes}, but does not provide alternative coherent equilibria that might offer better payoffs
to the players.

\begin{figure}
\centerline{\epsfig{file=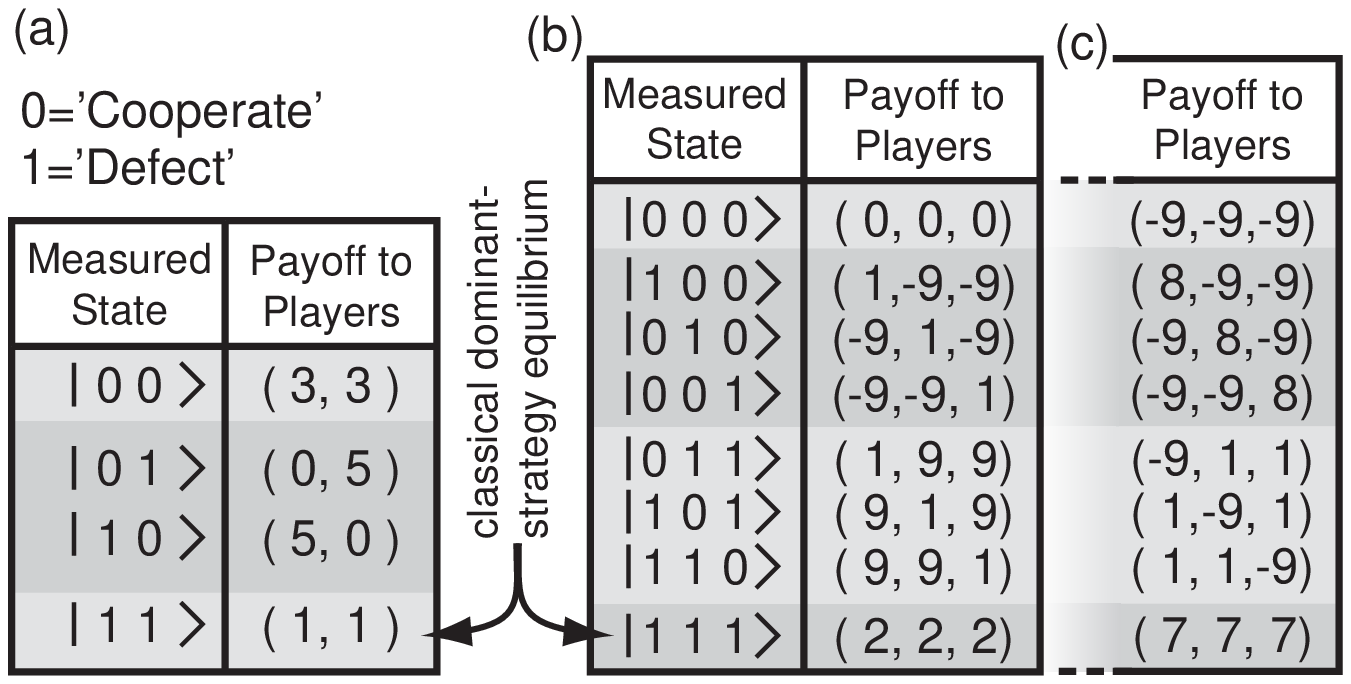,width=7.8cm}}
\vspace{0.2cm}
\caption{Games possessing a dominant-strategy equilibrium: (a) table defining the payoffs in
Prisoner's Dilemma. Either player reasons thus: `If my partner were to cooperate, my best action would
be to defect. If he were to defect, my best action is still to defect. Thus I have a dominant strategy:
``always defect.''\ ' (b) Table defining the payoffs for a three-player game. Classically, each player
has the dominant strategy `choose 1'. Consequently, each player's payoff is just 2 points (despite the
existence of strategy profiles, such as `choose 1 with probability 80\%', for which all the players have
greater expected payoffs). However in the quantum game, 
radically superior new coherent equilibria arise. (c) A game where quantum players do {\em less} well than their classical counterparts.}
\label{figure2}
\end{figure}

To investigate the multi-player case, we quantize the game
of Figure 2(b). We find that coherent equilibria {\em do} occur. The classically inevitable outcome, now
written as $({\hat F},{\hat F},{\hat F})$, becomes a Nash equilibrium -- but other, radically superior
equilibria emerge. For example, the profile
$s = ({\hat I}, {1\over\sqrt{2}}({\hat \sigma}_x+{\hat \sigma}_z),{\hat \sigma}_x)$, with expected payoffs
$(5,9,5)$, is a Nash equilibrium (and is {\em strict} for
players A and C: any unilateral deviation necessarily leads to {\em reduction} in their expected payoffs). Note that there is no in-principle difficultly with the asymmetry \cite{symmdilemma} of the profile, since in this game we are allowing free classical communication between players.
The proof that this profile is a Nash equilibrium runs as follows.

Let
$
\ket{\psi} =( \hat{I} \otimes 
	\frac{i}{\sqrt{2}}({\hat \sigma}_x + {\hat \sigma}_z)  \otimes \hat{I}) \hat{J} \ket{000}
$
be the state after the actions of players A and B, and
suppose that player C applies a general open quantum operation ${\cal R}$, i.e.
a completely positive,
trace-preserving map on density operators.  By the Kraus representation
theorem \cite{kraus}, we can write
${\cal R}(\rho)=\sum_k {\hat A}_k \rho {\hat A}_k^\dagger$,
under the restriction $\sum_k {\hat A}_k^\dagger {\hat A}_k = {\hat I}$.  We
may think of this expansion as representing a $k$-outcome
measurement, where it is allowed to perform unitary operations
conditioned on the outcome of the measurement.
The state-change corresponding to outcome $k$
is given by
$
\ket{\psi} \mapsto (\bra{\psi} {\hat A}_k^\dagger {\hat A}_k \ket{\psi})^{-{1\over 2}}{\hat A}_k \ket{\psi}.
$
Since player C only applies local operations, the most general
$
{\hat A}_k = {\hat I} \otimes {\hat I} \otimes \left( \begin{array}{cc}
a & b \\
c & d
\end{array} \right)
$. But it is then simple to show, by applying this ${\hat A}_k$ followed by the gate ${\hat J}^\dagger$,
that player C's expected payoff is maximized only if ${\hat A}_k \propto {\hat \sigma}_x$. Therefore, the
only strategy for player C which maximizes her expected payoff for every one of
her measurement outcomes is, up to global phase, $\sigma_x$. By repeating similar arguments for players A 
and B, we verify that
$s$ is indeed a Nash equilibrium for the full quantum strategy space $S_{TCP}$.

We have seen that superior quantum coherent equilibria occur in some games (the 3 player Dilemma and the 4 player Minority Game), but are absent in others (the 3 player Minority, and any maximally entangled fair 2 player game). But do quantum players always fare at least as well as their classical counterparts? No. Figure 2 (c) is the payoff table for a game with a very high-performing dominant strategy - since all other outcomes have much lower total payoffs, this classically inevitable outcome is optimal.
But in the quantized game this profile, now
written as $({\hat F},{\hat F},{\hat F})$, is no longer even a Nash equilibrium: any player can unilaterally improve her payoff by switching to $s=i\sigma_y$, with severe consequences for the other players. Therefore, any equilibria in the quantum game will be inferior to the classical equilibrium.

To conclude, we have performed the first investigation of multi-player quantum games, finding that such games can exhibit forms of `pure' quantum equilibrium which have no analogue in
classical games, or even in two-player quantum games. In the Minority Game, we found that the players were able to exploit entanglement to overcome the frustration in the classical variant, and so play the game `perfectly'. More dramatically, in our Dilemma game the quantum players can escape the classical Dilemma entirely: they can play cooperatively knowing that no player can successfully `defect' against the others. In this respect quantum entanglement fulfills the role of a contract. Our observation of these purely coherent equilibria
paves the way for an investigation into iterated quantum games \cite{NeilsPaper}.

SB was initially attracted to this topic by several inspiring discussions with Neil Johnson, who conjectured that new
forms of equilibria would result from the complexity of the $N>2$ player minority game. We also acknowledge helpful
discussions with Art Pittenge, Vlatko Vedral and Julia Kempe. SB
and PH are supported by EPSRC and the Rhodes Trust, respectively. The authors would also like to acknowledge funding from the EU QAIP
project under contract EC-IST-1999 11234.


\begin{references}
\bibitem{whySigmaX}
The most general
quantum bit-flip operator (up to a global phase) is
$\cos(\theta){\hat \sigma}_x+\sin(\theta){\hat \sigma}_y$. Our ${\hat F}=\sigma_x$, but any other
choice is equally valid, and would simply result in a trivial rotation of the features which we 
discover.
\bibitem{whatIsMax} 
Since we are interested in purely multipartite entanglement, we call
a (pure) state maximally entangled if it is equivalent via local unitary
operations to the GHZ-type state, 
$1/\sqrt{2}(\ket{00 \cdots 0} + \ket{11 \cdots 1})$.
\bibitem{stillGeneral} Any other ${\hat J}$ meeting these conditions would be equivalent, via local
unitary operations, to our ${\hat J}$. Therefore the resulting game equilibria would be equivalent, via the
same operators, to the equilibria we discover.
\bibitem{whyProbSU2destEnt} 
If the action is not drawn from $S_U$ then the resulting state
is necessarily not equivalent via local unitary operations to a GHZ state.
Such a state is therefore no longer `maximally entangled' in our sense.
\bibitem{manyNash} Interestingly, the complexity of the four player game is such that several Nash
equilibria occur. Players may use criteria, such as maximum projection into the subspace
of `classical' moves $\{{\hat I},{\hat F}\}$, to establish a focal point, but this becomes a psychological 
question.
\bibitem{additClassRes}
If we generate a pair of random classical bits, and send duplicates of these bits to
each player, then Nash equilibria exist wherein the players use these bits to decide
which player should be in the minority.
\bibitem{symmdilemma} There is a focal symmetric strategy profile in this game, where each player adopts $s={1\over \sqrt2}(I+i\sigma_y)$. However this profile is not a Nash equilibrium unless the players are actually constrained to unitary moves. One could presumably construct a similar game wherein this strategy profile does form a full Nash equilibrium - e.g. the game obtained by replacing the payoff columns in Fig 2(b) by $(8,8,8)$, $(1,-9,-9)$, $(-9,1,-9)$, $(-9,-9,1)$, $(0,4,4)$, $(4,0,4)$, $(4,4,0)$, $(1,1,1)$ (top to bottom) looks promising in this respect.
\end{references}
\end{document}